\documentclass[xdvi,twoside]{article}
\usepackage{fleqn,espcrc2,graphicx}

\newcommand{\AmS}{{\protect\the\textfont2
  A\kern-.1667em\lower.5ex\hbox{M}\kern-.125emS}}

\title{Moments of Structure Functions in Full QCD\thanks{Talk presented 
       by J. W. Negele. Work supported in part by the U.S. Department
       of Energy (DOE) under cooperative research agreement \#
       DE-FC02-94ER 40818. }}
\author{D. Dolgov, R. Brower, S. Capitani, J. W. Negele, A. Pochinsky,
D. Renner \\
\vspace{.1cm}
Center for Theoretical Physics, MIT, 77 Massachusetts Ave.,  Cambridge, MA 02139, USA \\
\vspace{.2cm}
N. Eicker, T. Lippert,   K. Schilling \\
\vspace{.1cm}
Department of Physics, University of Wuppertal, D-42097 Wuppertal, Germany\\
\vspace{.2cm}
R. G. Edwards,$^a$ U. M. Heller$^b$ \\
\vspace{.1cm} 
$^a$Jefferson Lab, 12000 Jefferson Avenue, MS 12H2,
Newport News, Virginia 23606, USA \\
\vspace{.1cm}
$^b$CSIT, Florida State University, Tallahassee FL 32306 USA}

\begin{document}
\begin{abstract}
Moments of the quark density distribution, moments of the quark helicity
distribution, and the tensor charge are calculated in full QCD.
Calculations of matrix elements of operators from the operator product 
expansion have been performed on $16^3 \times 32$ lattices for Wilson 
fermions at $\beta = 5.6$ using
configurations from the SESAM collaboration and at $\beta = 5.5$ using
configurations from SCRI. One-loop perturbative renormalization
corrections are included. Selected results are compared with corresponding
quenched calculations and with calculations using cooled configurations.
\end{abstract}

\maketitle

\section{INTRODUCTION}
Given the
detailed experimental knowledge of the light cone
distributions of quarks and gluons in the nucleon, it is of interest
to use lattice QCD both to calculate the quark and gluon
structure of the nucleon from first principles and 
to reveal the underlying mechanisms giving rise to this
structure.  Using the
operator product expansion, it is possible to calculate
moments of quark distributions, and we report here the first
calculations in full QCD \cite{dolgov-thesis}.
We also  compare full QCD results with quenched QCD and with
configurations that have been cooled to remove all the gluon
contributions except for those of instantons.  

\section{DEFINITIONS}
The moments of the spin-independent structure functions $F_1(x, Q^2)$,
$F_2(x, Q^2)$ and the spin-dependent structure functions $g_1(x, Q^2)$, 
$g_2(x, Q^2)$ are related 
to products of Wilson coefficients $C_n(Q^2/ \mu^2)$ times hadronic matrix
elements:
 
\smallskip
\noindent$
\int^1_0dx\,x^n \,F_1(x,Q^2) =\frac{1}{2}
C_{n+1}^{v}(\frac{Q^2}{\mu^2})\, \langle x^n q\rangle(\mu) \nonumber \\
\int^1_0dx\,x^n \,F_2(x,Q^2) = 
C_{n+2}^{v}(\frac{Q^2}{\mu^2})\, \langle x^{n+1} q\rangle (\mu) \nonumber \\
\int^1_0dx\,x^n\,g_1(x, Q^2) = \frac{1}{4} 
C_n^{a}(\frac{Q^2}{\mu^2})\, 2 \langle x^n \Delta q\rangle (\mu) \nonumber \\
\int^1_0dx\,x^n\,g_2(x, Q^2) = \frac{1}{4} \frac{n}{n+1}
\big[C_n^{d}(\frac{Q^2}{\mu^2}) \, d_n(\mu) \nonumber$ \\
\hspace*{.7in}$\quad \quad \quad \quad \quad  -C_n^{a}(\frac{Q^2}{\mu^2}) \, 2 \langle x^n 
\Delta q\rangle \big] . \label{dis-momsum}$

\smallskip

These matrix elements, denoted $\langle x^n q \rangle$ (which equals
$v_{n+1}^{(q)}$ in the notation of ref.\cite{qcdsf}),
$\langle x^n\Delta q \rangle$ (which equals $\frac{1}{2}a_{n}^{(q)}$
\cite{qcdsf}) and  $d$, as well as the moments of the transversity distribution
$h(x,Q^2)$,    $\langle x^n \delta q \rangle$, are related to expectation
values of the following operators in the proton ground state:

\smallskip
\noindent $
{2 \langle x^{n-1} q \rangle P_{\mu_1} \cdots P_{\mu_n}} $\\
\hspace*{.3in}$= \frac{1}{2} \langle PS | \left(\frac{i}{2}\right)^{n-1} 
\bar{\psi}\gamma_{\{\mu_1}{\stackrel{\,\leftrightarrow}{D}}_{\mu_2} \cdots
{\stackrel{\,\leftrightarrow}{D}}_{\mu_n\}}\psi | PS\rangle$ \\

\vspace*{-.1in}\noindent
${\frac{2}{n} \langle x^n \Delta q \rangle S_{\{\sigma}P_{\mu_1}\cdots P_{\mu_n\}}} $\\
\hspace*{.3in}$= -\langle PS | \left(\frac{i}{2}\right)^{n} \bar{\psi}
\gamma_5\gamma_{\{\sigma}{\stackrel{\,\leftrightarrow}{D}}_{\mu_1} \cdots
{\stackrel{\,\leftrightarrow}{D}}_{\mu_n\}}\psi | PS\rangle $ \\

\vspace*{-.1in}\noindent
$\frac{1}{n} {d_n S_{[\sigma}P_{\{\mu_1]}\cdots P_{\mu_n\}}}  $\\
\hspace*{.3in}$= -\langle PS | \left(\frac{i}{2}\right)^{n} \bar{\psi}
\gamma_5\gamma_{[\sigma}{\stackrel{\,\leftrightarrow}{D}}_{\{\mu_1]} \cdots
{\stackrel{\,\leftrightarrow}{D}}_{\mu_n\}}\psi | PS\rangle $ \\

\vspace*{-.1in}\noindent
$\frac{1}{m_N} { \langle x^{n} \delta q \rangle
S_{[\mu}P_{\{\nu]}P_{\mu_1}\cdots P_{\mu_n\}}}  $\\
\hspace*{.3in}$= \langle PS | \left(\frac{i}{2}\right)^{n} \bar{\psi}
\gamma_5\sigma_{\mu\{\nu }{\stackrel{\,\leftrightarrow}{D}}_{\mu_1} \cdots
{\stackrel{\,\leftrightarrow}{D}}_{\mu_n\}}\psi | PS\rangle . $

\smallskip
\smallskip
In the parton model, these matrix elements correspond to the following 
moments of quark longitudinal and transverse distributions:

\smallskip
\noindent $x^nq \, \, \, \,  
\equiv  \int_0^1dx\,x^n\,(q_\downarrow (x)+q_\uparrow (x)) \\ 
x^n \Delta q \equiv  \int_0^1dx\,x^n\,(q_\downarrow (x)-q_\uparrow (x))
\nonumber \\
x^n \delta q \, \, \equiv  \int_0^1dx\,x^n\,(q_\top (x)-q_\bot (x)) . $

\smallskip
\section{SOURCES}
\begin{figure}[t]
\includegraphics[angle=-90,width=18pc]{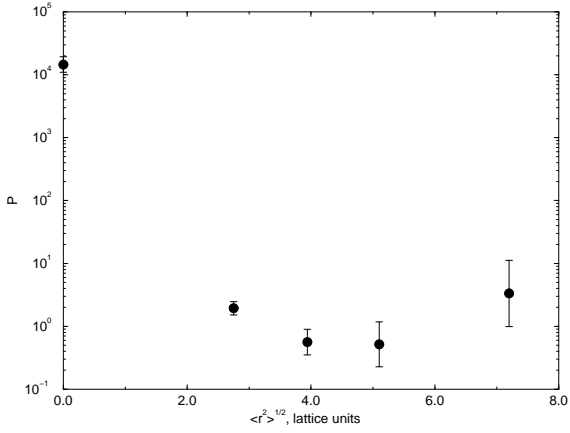}
\vspace{-28pt}
\caption{Contamination by excited states, $P$,  as a function of the
source  RMS radius}
\label{fig-smsm-twopt-vs-rms-ABB-log}
\end{figure}
Connected diagrams are calculated using sequential
propagators generated by the upper two components of the
nucleon source $J^{\alpha} =
u_a^{\alpha}u_b^{\beta}(C\gamma_5)_{\beta,\beta'}
d_c^{\beta'}\epsilon^{abc}$. The overlap with the physical
proton ground state was optimized using Wuppertal
smearing \cite{sources} to minimize the contamination $P = {\sum_{n \neq
0}|\langle J|n\rangle |^2}/ {|\langle J|0\rangle |^2}$.
Figure~\ref{fig-smsm-twopt-vs-rms-ABB-log}
shows that varying the smearing reduced $P$ by over 4
orders of magnitude, yielding an overlap with the physical
ground state of approximately 70\%. Dirichlet boundary
conditions were used for quarks in the t-direction.

\section{OPERATORS AND
PERTURBATIVE RENORMALIZATION}
\begin{center}
\begin{small}
\begin{table}
\caption{Operators used to measure moments of quark distributions}
\label{tab-operators}
\begin{tabular}{llrl}
\hline
& H(4) mix&$\vec{p}$&lattice operator\\
\hline
$xq_c^{(a)}$ & {\bf 6}$_3^+$ \, no & $\neq 0$  & $\bar{q} \gamma_{\{1} 
{\stackrel{\,\leftrightarrow}{D}}_{4\}} {q}$  \\
$xq_c^{(b)}$ & {\bf 3}$_1^+$ \, no & 0& $\bar{q} \gamma_{4} 
{\stackrel{\,\leftrightarrow}{D}}_{4} {q}$ \\
& & \multicolumn{2}{c}{ $~~~- \frac{1}{3} \sum_{i=1}^3
 \bar{q} \gamma_{i} {\stackrel{\,\leftrightarrow}{D}}_{i} {q}$}  \\
$x^2q_c$     & {\bf 8}$_1^-$ \, yes & $\neq 0$ & $\bar{q} \gamma_{\{1} 
{\stackrel{\,\leftrightarrow}{D}}_{1} 
{\stackrel{\,\leftrightarrow}{D}}_{4\}} {q}$ \\
& & \multicolumn{2}{c}{ $~~~- \frac{1}{2} \sum_{i=2}^3
\bar{q}\gamma_{\{i} {\stackrel{\,\leftrightarrow}{D}}_{i} 
{\stackrel{\,\leftrightarrow}{D}}_{4\}}{q}$} \\
$x^3q_c$     & {\bf 2}$_1^+$ \, no$^*$ & $\neq 0$ & $\bar{q} \gamma_{\{1} 
{\stackrel{\,\leftrightarrow}{D}}_{1} 
{\stackrel{\,\leftrightarrow}{D}}_{4} 
{\stackrel{\,\leftrightarrow}{D}}_{4\}} {q}$ \\
& & \multicolumn{2}{c}{ $~~~+ \bar{q} \gamma_{\{2} {\stackrel{\,\leftrightarrow}{D}}_{2} 
{\stackrel{\,\leftrightarrow}{D}}_{3} 
{\stackrel{\,\leftrightarrow}{D}}_{3\}} {q}$} \\
& & \multicolumn{2}{c}{ $~~~- (\,3\,\leftrightarrow\,4\,)$} \\
\hline
$\Delta q_c$     & {\bf 4}$_4^+$ \, no & 0  & $\bar{q} 
\gamma^{5} \gamma_{3} {q}$  \\
$x\Delta q_c^{(a)}$ & {\bf 6}$_3^-$  \, no & $\neq 0$ & $\bar{q} 
\gamma^5 \gamma_{\{1} {\stackrel{\,\leftrightarrow}{D}}_{3\}} {q}$ \\
$x\Delta q_c^{(b)}$ & {\bf 6}$_3^-$  \, no & 0 & $\bar{q} \gamma^5 
\gamma_{\{3} {\stackrel{\,\leftrightarrow}{D}}_{4\}} {q}$ \\
$x^2\Delta q_c$     & {\bf 4}$_2^+$  \, no & $\neq 0$ & $\bar{q} 
\gamma^5 \gamma_{\{1} {\stackrel{\,\leftrightarrow}{D}}_{3} 
{\stackrel{\,\leftrightarrow}{D}}_{4\}} {q}$ \\
\hline
$\delta q_c$     & {\bf 6}$_1^+$  \, no & 0  & $\bar{q} \gamma^{5} \sigma_{34} 
{q}$  \\
$x\delta q_c$    & {\bf 8}$_1^-$ \, no & $\neq 0$   & $\bar{q} \gamma^{5} 
\sigma_{3\{4} {\stackrel{\,\leftrightarrow}{D}}_{1\}}{q}$  \\
$d_1$     & {\bf 6}$_1^+$  \, no$^{**}$ & 0  & $\bar{q} \gamma^5 \gamma_{[3} 
{\stackrel{\,\leftrightarrow}{D}}_{4]}{q}$ \\
$d_2$     & {\bf 8}$_1^-$  \, no$^{**}$ & $\neq 0$ & $\bar{q} \gamma^5 
\gamma_{[1} {\stackrel{\,\leftrightarrow}{D}}_{\{3]} 
{\stackrel{\,\leftrightarrow}{D}}_{4\}} {q}$ \\
\hline
\end{tabular}
\end{table} 
\end{small}
\end{center}
\begin{table}[!hbt]
\caption{Perturbative renormalization constants}
\label{tab-renorm1}
\begin{center}
\begin{small}
\setlength{\tabcolsep}{5pt}
\begin{tabular}{lcrcll}
\hline
& $\gamma$ & $B^{LATT}$ & $B^{{\overline{MS}}}$ & \multicolumn{2}{c}{$Z$} \\
& & & & ${}_{\beta=6.0}$&${}_{\beta=5.6}$ \\
\hline
$xq^{(a)}$         &     $\frac{8}{3}$  &   $-3.16486$  &      $-\frac{40}{9}$
   & $0.989$& $0.988$ \\
$xq^{(b)}$         &     $\frac{8}{3}$  &   $-1.88259$  &      $-\frac{40}{9}$
   & $0.978$& $0.977$ \\
$x^2q$             &    $\frac{25}{6}$  &  $-19.57184$  &      $-\frac{67}{9}$
   & $1.102$& $1.110$ \\
$x^3q$             &  $\frac{157}{30}$  &  $-35.35192$  &  $-\frac{2216}{225}$
   & $1.215$& $1.231$ \\
$\Delta q$         &               $0$  &   $15.79628$  &                  $0$
   & $0.867$& $0.857$ \\
$x\Delta q^{(a)}$  &     $\frac{8}{3}$  &   $-4.09933$  &      $-\frac{40}{9}$
   & $0.997$& $0.997$ \\
$x\Delta q^{(b)}$  &     $\frac{8}{3}$  &   $-4.09933$  &      $-\frac{40}{9}$
   & $0.997$& $0.997$ \\
$x^2\Delta q$      &    $\frac{25}{6}$  &  $-19.56159$  &      $-\frac{67}{9}$
   & $1.102$& $1.110$ \\
$\delta q$         &               $1$  &   $16.01808$  &                 $-1$
   & $0.856$& $0.846$ \\
$x\delta q$        &               $3$  &   $-4.47754$  &                 $-5$
   & $0.996$& $0.995$ \\
$d_1$              &               $0$  &  $0.36500$  &                  $0$
   & $0.997$& $0.997$ \\
$d_2$              &     $\frac{7}{6}$  &  $-15.67745$  &     $-\frac{35}{18}$
   & $1.116$& $1.124$ \\
\hline
\end{tabular}
\end{small}
\end{center}
\end{table}
\begin{figure}[hbt]
\includegraphics[angle=-90,width=18pc]{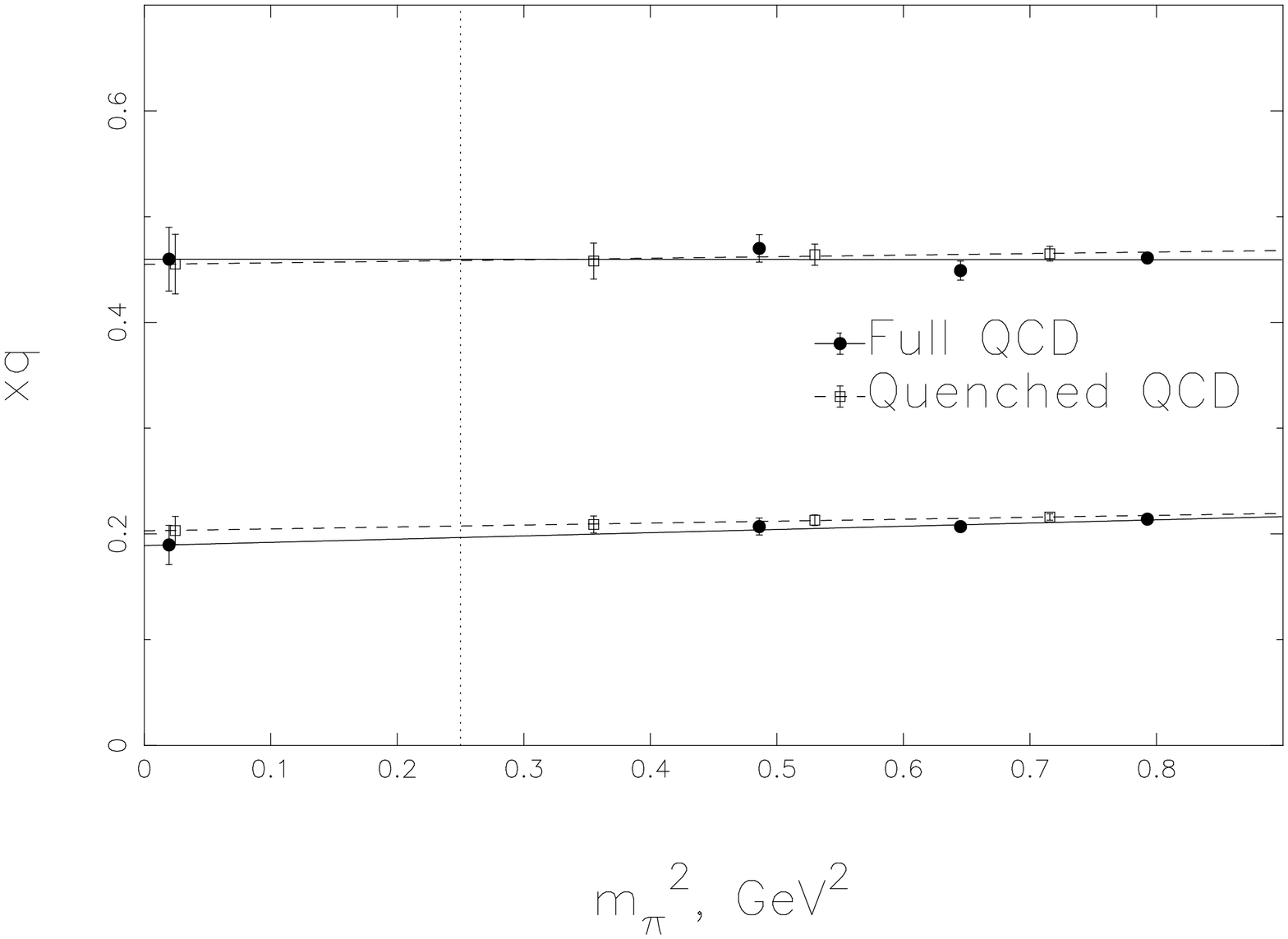}

\includegraphics[angle=-90,width=18pc]{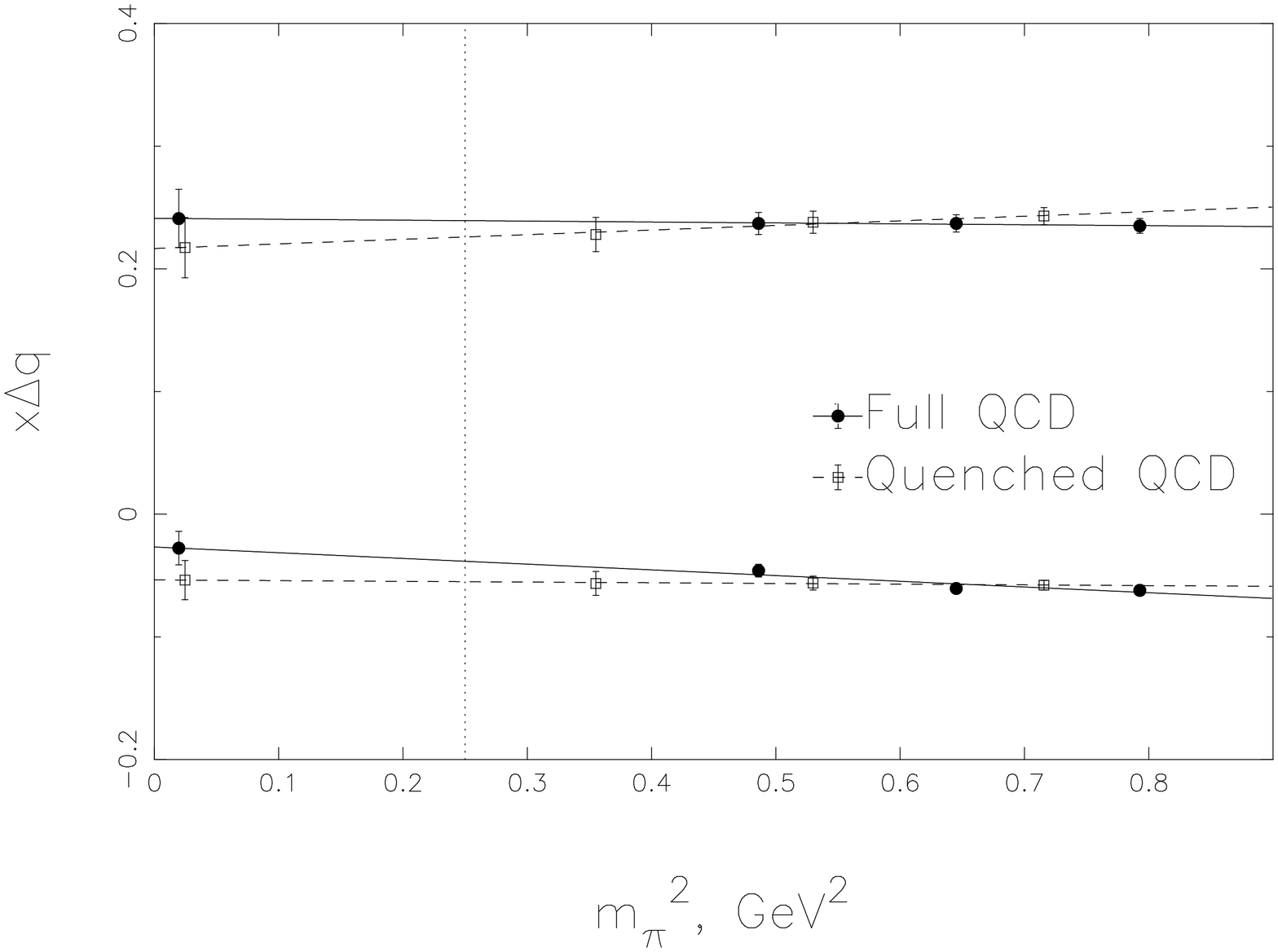}
\vspace{-28pt}
\caption{Comparison of chiral extrapolations of full and quenched
  calculations of $\langle x q \rangle$ and  $\langle x \Delta q
   \rangle$  showing agreement within statistical errors   }
\label{fig-full}
\end{figure}
\begin{figure}[hbt]
\includegraphics[angle=-90,width=18pc]{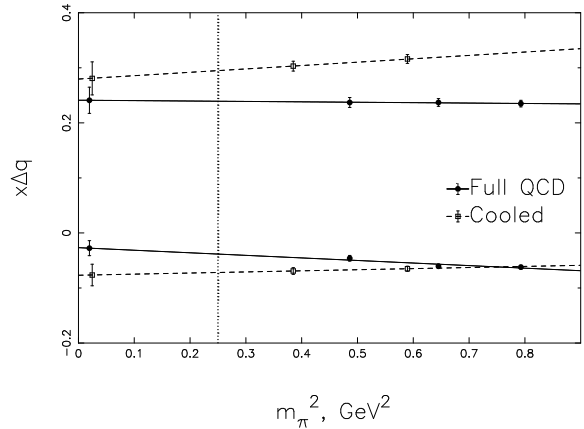}
\vspace{-28pt}
\caption{Comparison of chiral extrapolations of full and cooled
  calculations of $\langle x \Delta q \rangle$ showing the extent to
   which instantons reproduce the full result}
  
\label{fig-full-cooled}
\end{figure}
The continuum operators defined above are approximated on a
discrete cartesian lattice using representations of the
hypercubic group that eliminate operator mixing where
possible and minimize the number of non-zero
components of the
nucleon momentum.  The
operators we have used are shown in Table~\ref{tab-operators}, where we
have indicated whether the spatial momentum components
are non-zero and whether mixing occurs. (Note, no$^*$
indicates a case in which mixing could exist in general but
vanishes perturbatively for Wilson or overlap fermions and
no$^{**}$ indicates perturbative mixing with {\it lower}
dimension operators for Wilson fermions but no mixing for
overlap fermions.)

The perturbative renormalization coefficients we have
calculated and used in this work are tabulated in
Table~\ref{tab-renorm1} \cite{pert-renorm}. 
The factor to convert lattice results to the continuum $ \overline{MS}$
scheme is $Z(g_0^2 = 6/\beta) = 1 - \frac{g_0^2}{16 \pi^2} \frac{4}{3}
(B^{LATT} - B^{\overline{MS}})$.  

\begin{table*}[hbt]
\caption{Comparison of our full QCD and quenched results with other
  lattice calculations and phenomenology at  $4$ GeV in $\overline{MS}$}
\label{tab-summary}
\begin{center}
\begin{small}
\begin{tabular}{lrrrrrr}
\hline
& QCDSF & QCDSF & Wuppertal &Quenched &Full QCD 
& \multicolumn{1}{c}{Phenomenology} \\
&  & ($a=0$) &  &  & (3 pts) & \multicolumn{1}{c}{($q_{val}$)} \\  
\hline
$ x u_c $      & $0.452(26)$ &   &   & $0.454(29)$ & $0.459(29)$ & $0.284$ \\
$ x d_c $      & $0.189(12)$ &   &   & $0.203(14)$ & $0.190(17)$ & $0.104$ \\
$xu_c-xd_c$    & $0.263(17)$ &   &   & $0.251(18)$ & $0.269(23)$ & $0.180$ \\ 
$ x^2 u_c $    & $0.104(20)$ &   &   & $0.119(61)$ & $0.176(63)$ & $0.083$ \\
$ x^2 d_c $    & $0.037(10)$ &   &   & $0.029(32)$ & $0.0314(303)$ & $0.025$ \\
$ x^3 u_c $    & $0.022(11)$ &   &   & $0.037(36)$ & $0.0685(392)$ & $0.032$ \\
$ x^3 d_c $    & $-0.001(7)$ &   &   & $0.009(18)$ & $-0.00989(1529)$ & $0.008$ \\
$ \Delta u_c $ & $0.830(70)$ & $0.889(29)$ & $0.816(20)$
& $0.888(80)$ & $0.860(69)$ & $0.918$ \\
$ \Delta d_c $ & $-0.244(22)$ & $-0.236(27)$ & $-0.237(9)$
& $-0.241(58)$ & $-0.171(43)$ & $-0.339$ \\
$ \Delta u_c-\Delta d_c $ & $1.074(90)$ & $1.14(3)$ & $1.053(27)$  
& $1.129(98)$ & $1.031(81)$ & $1.257$ \\
$ x\Delta u_c $ & $0.198(8)$ &      &  & $0.215(25)$ & $0.242(22)$ & $0.150$ \\
$ x\Delta d_c $ & $-0.048(3)$ &      &  & $-0.054(16)$ & $-0.0290(129)$ & 
 $-0.055$ \\
$ x^2 \Delta u_c $ & $0.087(14)$ &      &  & $0.027(60)$ & $0.116(42)$ & $0.050$ \\
$ x^2 \Delta d_c $ & $-0.025(6)$ &      &  & $-0.003(25)$ & $0.00142(2515)$ & $0.016$ \\
$ \delta u_c $ & $0.93(3)$ & $0.980(30)$ &   & $1.01(8)$ & $0.963(59)$ &  \\
$ \delta d_c $ & $-0.20(2)$ & $-0.234(17)$ &   & $-0.20(5)$ & $-0.202(36)$ &  \\
$ d_2^u $ & $-0.206(18)$ &     &   & $-0.233(86)$ & $-0.228(81)$ &  \\
$ d_2^d $ & $-0.035(6)$ &     &   & $0.040(31)$ & $0.0765(310)$ &  \\
\hline
\end{tabular}
\end{small}
\end{center}
\end{table*}

\section{RESULTS}
 
The moments listed in Table 1 were calculated \cite{dolgov-thesis} 
on $16^3 \times 32$ lattices for Wilson 
fermions in full QCD at $\beta = 5.6$ using 200 SESAM configurations
at each of 4 $\kappa 's$ and at $\beta = 5.5$ using 100 SCRI
configurations at 3 $\kappa 's$. 
They were also calculated with two sets of 100 full
QCD configurations cooled with 50 cooling steps and in
quenched  QCD at $\beta = 6.0$ using 200 configurations at each of 3
$\kappa 's$.  Typical chiral extrapolations for operators calculated
with nucleon momentum equal to zero are shown in Figure~\ref{fig-full}
for full and quenched
calculations of $\langle x q \rangle$ and  $\langle x \Delta q
\rangle$, showing agreement within statistical errors. To avoid
finite volume errors at the 
lightest quark mass, the SESAM \cite{sesam} results were extrapolated using the
three heaviest quark masses. Table~3 shows a major result of our work,
that there is complete
agreement within statistics between full and quenched
results. Statistics with the SCRI configurations \cite{scri} 
are not yet adequate
to present extrapolations in the coupling constant.

Typical
chiral extrapolations for cooled configurations are compared with the
corresponding uncooled full QCD calculations in
Figure~\ref{fig-full-cooled}. 
This qualitative agreement between cooled and uncooled results occurs at
light quark mass for
all the twist-2 matrix elements we calculated  and demonstrates the
degree to which the instanton content of the configurations 
and their associated zero modes dominate light hadron structure 
\cite{instantons}.

Comparison in Table~3 of our quenched results with those of the QCDSF
collaboration \cite {qcdsf} and the full QCD results with those of Wuppertal
\cite{sesam} shows complete consistency. Since the phenomenological
quantity $q_{val}$ \cite{phenomenology} does not correspond precisely
to the connected diagrams we calculate, the most meaningful comparison
is with $u - d$ differences. The two most physically significant
discrepancies arising from this table are the fact that the difference in
first moment is $ \sim $ 0.25 - 0.29 on the lattice and 0.18
experimentally and the axial charge is $ \sim $ 1.0 - 1.1 on the lattice
and 1.26 experimentally. We have clearly shown that these
discrepancies do not arise from quenching and believe both indicate
inadequate treatment of the pion cloud of the nucleon due to the small
physical volume and heavy quark mass. 


\end{document}